%% file: main.tex
\documentclass{article}
\usepackage{spconf,amsmath,graphicx}
\usepackage{MnSymbol}
\usepackage{amsthm}
\usepackage{enumitem}
\usepackage{amsfonts}
\usepackage{algorithm}
\usepackage{algorithmicx,algpseudocode}

\usepackage[tight,footnotesize]{subfigure}
\usepackage{multirow}
\usepackage{pbox}
\usepackage{url}
\usepackage{comment}
\usepackage{color}

\input{macros}

\title{Perceptually inspired weighted MSE optimization using irregularity-aware graph Fourier transform}
\name{
    Keng-Shih~Lu$^{*}$, Antonio~Ortega$^{*}$, Debargha~Mukherjee$^{\dagger}$, and Yue~Chen$^{\dagger}$
}
\address{$^{*}$University of Southern California, Los Angeles, USA \\ $^{\dagger}$Google LLC, Mountain View, USA}

\begin{document}
\ninept
\maketitle
\begin{abstract}
In image and video coding applications, distortion has been traditionally measured using mean square error (MSE), which suggests the use of orthogonal transforms, such as the discrete cosine transform (DCT). Perceptual metrics such as Structural Similarity (SSIM) are typically used after encoding, but not tied to the encoding process. In this paper, we consider an alternative framework where the goal is to optimize a \emph{weighted MSE} metric, where different weights can be assigned to each pixel so as to reflect  their relative importance in terms of  perceptual image quality. For this purpose, we propose a novel transform coding scheme based on irregularity-aware graph Fourier transform (IAGFT), where the induced IAGFT is orthogonal, but the orthogonality is defined with respect to an inner product corresponding to the weighted MSE. We propose to use weights derived  from local variances of the input image, such that the weighted MSE aligns with SSIM. In this way, the associated IAGFT can achieve a coding efficiency improvement in SSIM with respect to conventional transform coding based on DCT. Our experimental results show a compression gain in terms of multi-scale SSIM on test images. 
\end{abstract}
\vspace{-.2cm}
\section{Introduction}
\label{sec:intro}
% image/video coding transform coding and GSP
Most image and video compression systems make use of transform coding, where correlations among pixels can be exploited in order to concentrate most signal energy in a few frequencies. The widely used discrete cosine transform (DCT) \cite{strang1997wavelets} has been shown to achieve optimal decorrelation when when pixel data can be modeled as a Markov random field with high correlation \cite{zhang2013analyzing}. In recent years, graph signal processing (GSP) tools \cite{shuman2013emergin,sandryhaila2013discrete,ortega2018graph} have been applied to image and video coding to enhance coding efficiency \cite{egilmez2019graph-based,hu2016graph-based,cheung2018graph}.
GSP is a framework that extends conventional signal processing tools to signals supported on graphs, where data points and their inter-connections are captured by graph vertices and edges. In fact, the widely used discrete cosine transform (DCT) \cite{strang1999discrete} and asymmetric discrete sine transform (ADST) \cite{han2012jointly} are graph Fourier transforms (GFT) of two particular line graphs. 

While mean square error (MSE) is commonly used as quality metric in many coding standards, it is well-known that MSE does not always reflect perceptual quality.
Therefore, it is important to incorporate a perceptually-driven metric into the coding optimization process. Based on such a metric, it would be possible to formulate a bit allocation problem with the goal of spending more bits on image regions that are perceptually more sensitive to noise. In the literature, this problem is typically addressed by designing quantization strategies. For example, JPEG quantization tables can be designed based on human visual system (HVS) \cite{wang2001designing}, while JPEG-2000  adopts a visual masking technique \cite{zeng2000point-wise} that exploits self-contrast masking and neighborhood masking, leading to adaptive quantization of  wavelet coefficients without any overhead. Quantization parameter (QP) adjustment is a popular approach in video codecs such as HEVC \cite{sullivan2012overview}, in which QP is changed per block or per coding unit. 
In particular, H\"{o}ntsch et. al. \cite{hontsch2002adaptive} proposed an online update of the quantization step size, which is determined by a just-noticeable difference (JND) threshold based on local contrast sensitivity. Aside from the JND threshold, other key attributes to determine QP values would be region-of-interest \cite{meddeb2014region-of-interest} or statistics of transform coefficients \cite{seo2013rate}. 
%\cite{uz1993optimal}.... In \cite{hontsch2000locally}, locally-adaptive quantization scheme is proposed... In a technique called DCTune \cite{watson1993dctune}, a fixed quantization matrix is designed for a given image.
%Aside from QP adjustment, an alternative method is coefficient pruning. In \cite{cabrita2011perceptually}, the integration of QP adjustment and coefficient pruning is studied, where JND threshold is used for both method, leading to a compression gain in H.264 standard.

In this work, we propose a novel approach based on designing transforms with the goal of optimizing a weighted mean square error (WMSE), which allows us to adapt the perceptual quality pixel-wise instead of block-wise. We make use irregularity-aware graph Fourier transforms (IAGFTs) \cite{girault2018irregularity-aware}, generalized GFTs where orthogonality is defined with respect to an inner product such that distance between a signal and a noisy version corresponds to a WMSE instead of the MSE. 
A generalized Parseval's Theorem is then induced, in which the quantization error energy in the IAGFT transform domain is the same as the pixel domain WMSE. Based on the IAGFT, we design an image coding framework, where perceptual quality is characterized by choosing suitable weights for the WMSE. Under this framework, the overall perceptual quality of an image can be enhanced by weighting different pixels differently based on their perceptual importance, while the quantization step size is fixed for the entire image.
We consider a noise model, under which the WMSE weights are chosen to maximize the structural similarity (SSIM) \cite{wang2004image}.
%We demonstrate through an experiment that our framework fits into JPEG standard and provides a coding gain in SSIM.
We demonstrate experimentally the benefits of our framework by modifying a JPEG encoder to incorporate these novel transforms, showing coding gains in terms of multi-scale SSIM \cite{wang2003multiscale}.
In practice, the perceptual importance of pixels may vary spatially within an image block. Our method can take into account different weights for different pixels within a block, while existing QP adjustment methods can only adapt perceptual quality block-wise. To the best of our knowledge, perceptual coding scheme that is adaptive pixel-wise, and in transform domain, has not been studied in the literature. 

The rest of this paper is organized as follows. In Sec.~\ref{sec:preliminaries} we give a summary of graph signal processing and IAGFT. In Sec.~\ref{sec:perceptual_coding}, we propose a weight design for IAGFT that favors improved SSIM. Some properties of IAGFT basis are discussed in Sec.~\ref{subsec:bases}. In Sec.~\ref{sec:experiments}, we demonstrate of perceptually driven IAGFT through experimental results. Finally we conclude this paper in Sec.~\ref{sec:conclusion}.

\section{Preliminaries}
\label{sec:preliminaries}

\subsection{Graph Signal Processing}
\label{subsec:gsp}
We denote a weighted undirected graph as $\Gc=(\Vc,\Ec,\Wm)$, where $\Vc$ is the vertex set, $\Ec$ is the edge set of the graph, and $\Wm$ is the weighted adjacency matrix. In $\Wm$, the entry $w_{i,j}\geq 0$ represents the weight of edge $(i,j)\in\Ec$, and $w_{i,j}=0$ if $(i,j)\notin\Ec$. The graph Laplacian matrix is defined as $\Lm=\Dm-\Wm$, where $\Dm$ is the diagonal degree matrix with $d_{i,i}=\sum_{j=1}^n w_{i,j}$. The graph Laplacian can be viewed as a variation operator since the Laplacian quadratic form $\xv^\top\Lm\xv$ describes the variation of signal $\xv$ on the graph:
\begin{equation*}
\label{eq:lqf}
\xv^\top\Lm\xv = \sum_{(i,j)\in\Ec} w_{i,j}(x_i-x_j)^2
\end{equation*}

Following the definition in \cite{shuman2013emergin}, the (conventional) graph Fourier transform (GFT) is an orthogonal transform based on $\Um$,  the eigenvectors matrix of the graph Laplacian: $\Lm=\Um\Lambdam\Um^\top$. 
Based on this definition, the GFT basis functions $\uv_1$, $\dots$, $\uv_n$, i.e, the columns of $\Um$, are unit norm vectors corresponding to the increasing variations on the graph. In particular, if a random signal $\xv$ is modeled by a Gaussian Markov random field (GMRF) $\xv\sim \Nc(\zerov,\Lm^\dagger)$ \cite{rue2005gaussian}, then the GFT optimally decorrelates this signal. In fact, the 1D DCT is the GFT of a uniformly weighted line graph \cite{strang1999discrete}, which means that it optimally decorrelates pixel data that follows a uniform line graph model. Likewise, the 2D DCT is the GFT of a uniform grid graph.

\subsection{Irregularity-aware graph Fourier transform (IAGFT)}
\label{subsec:iagft}
The IAGFT \cite{girault2018irregularity-aware} is a generalization of the GFT, where the graph Fourier modes (i.e., GFT basis functions) are determined not only by the signal variation operator $\Lm$, but also by a positive definite matrix $\Qm$ 
that leads to a $\Qm-$inner product \cite{girault2018irregularity-aware}: $\left<\xv, \yv\right>_\Qm = \xv^\top\Qm\yv$, and therefore to a new definition of orthogonality: $\xv$ is orthogonal to $\yv$ if and only if $\xv^\top\Qm\yv =0$.
%One motivating application for IAGFT is in the context of sensor networks, where measured signals are independent of the distribution of the sensors, but signal energy in ordinary GFT fashion will highly depend on the sensor locations. 
Typically, $\Qm$ is chosen to be diagonal, so that the energy of signal $\xv$ is a weighted sum of its squared components: $\|\xv\|_\Qm^2=\sum_{i\in\Vc} q_i|x_i|^2$, with $\Qm=\text{diag}(q_1,\dots,q_n)$. 
The notion of generalized energy leads to a generalized GFT, i.e., the IAGFT:
\begin{definition}[Generalized graph Fourier modes]
\label{def:ggfmodes}
Given the Hilbert space defined by the $\Qm-$inner product and a graph variation operator $\Lm$, the set of $(\Lm,\Qm)-$graph Fourier modes is defined as the solution $\{\uv_k\}_k$ to the sequence of minimization problems: for increasing $K\in\{1,\dots,N\}$,
\begin{equation}
\label{eq:ggft_mode_problem}
    \underset{\uv_K}{\text{minimize}}\quad \uv_K^\top\Lm\uv_K \qquad \text{subject to} \quad \Um_K^\top \Qm \Um_K=\Id,
\end{equation}
where $\Um_K=(\uv_1, \dots, \uv_K)$.
\end{definition}
\begin{definition}[Irregularity-aware GFT]
\label{def:iagft}
Let $\Um$ be the matrix of $(\Lm, \Qm)-$graph Fourier modes, the $(\Lm,\Qm)-$GFT is $\Fm=\Um^\top\Qm$ and its inverse is $\Fm^{-1}=\Um$.
\end{definition}
In fact, \eqref{eq:ggft_mode_problem} can be written as a generalized Rayleigh quotient minimization, whose solution can be obtained efficiently through the generalized eigenvalue problem. Note that when $\Qm=\Id$, $\Fm$ reduces to conventional GFT as in Sec.~\ref{subsec:gsp}. One key property of the IAGFT is the \emph{generalized Parseval's theorem}: 
\begin{equation}
\label{eq:generalized_parseval}
    \left<\xv,\yv\right>_\Qm = \left<\hat{\xv},\hat{\yv}\right>_\Id,
\end{equation}
with $\hat{\xv}$ and $\hat{\yv}$ being the $(\Lm,\Qm)-$GFT of $\xv$ and $\yv$, respectively. Depending on the application, various choices of $\Qm$ may be used. Examples include diagonal matrices of node degrees and Voronoi cell areas (refer to \cite{girault2018irregularity-aware} for details).

\section{Perceptual coding with Weighted MSE}
\label{sec:perceptual_coding}
We focus on the weighted mean square error (WMSE) as an image quality metric, where different pixels are associated with different weights. First, in Sec.~\ref{subsec:coding_iagft}, we design a transform coding scheme that optimizes the WMSE, and analyze its error in pixel domain. We focus on the choice of perceptual quality inspired weights in Sec.~\ref{subsec:ssim_wmse}.

\subsection{Transform Coding with IAGFT}
\label{subsec:coding_iagft}

\begin{figure}
    \centering
    \subfigure[JPEG]{
    \includegraphics[scale=.33]{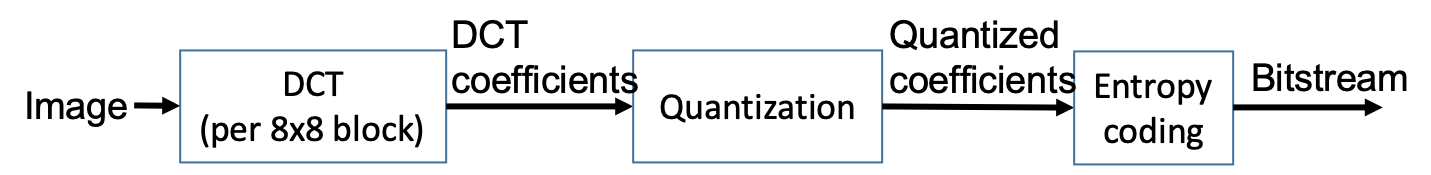}}
    \subfigure[Proposed]{
    \includegraphics[scale=.33]{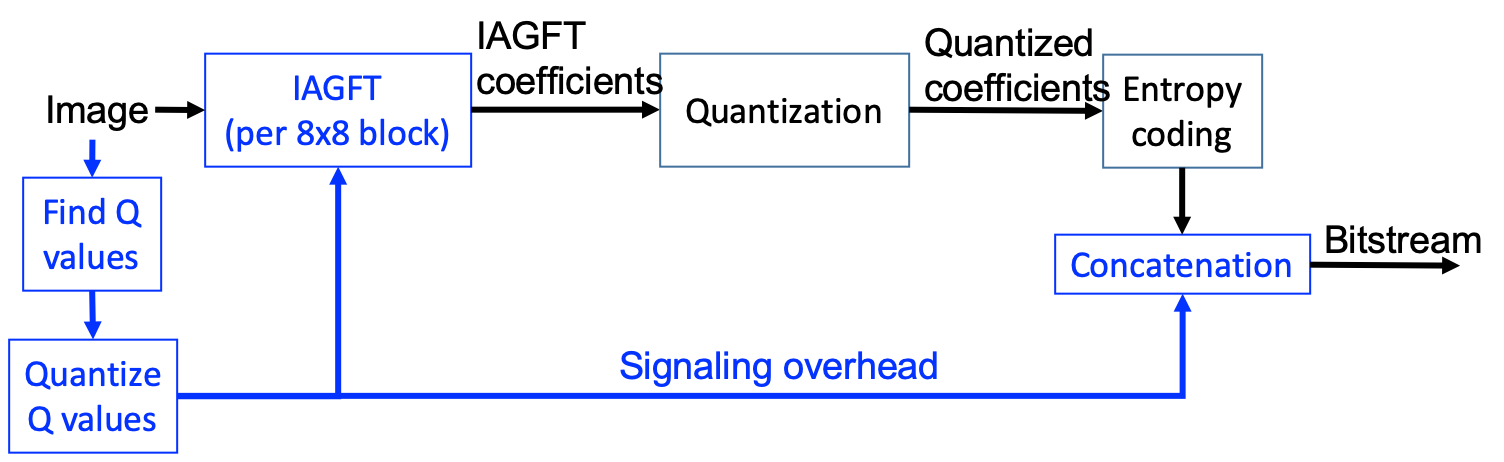}}
    \caption{Flow diagrams of JPEG and our proposed scheme. Blocks highlighted in blue are new components.}
    \label{fig:diagrams}
\end{figure}

We define the WMSE with weights $\qv\in\mathbb{R}^n$ (or, in short, $\qv$-MSE) between a distorted signal $\zv\in\mathbb{R}^n$ and its reference signal $\xv\in\mathbb{R}^n$  as
\begin{equation}
\label{eq:wmse}
    \text{WMSE}(\zv,\xv,\qv) := \frac{1}{n}\sum_{i=1}^n q_i(z_i-x_i)^2=\frac{1}{n}\left<\zv-\xv,\zv-\xv\right>_\Qm,
\end{equation}
where $\Qm=\text{diag}(\qv)$. When $\qv=\onev$, i.e., $\Qm=\Id$, the WMSE reduces to the conventional MSE. We note that the right hand side of \eqref{eq:wmse} is a $\Qm$-inner product, so the generalized Parseval's Theorem gives
\begin{equation*}
\label{eq:wmse_q}
    \text{WMSE}(\zv,\xv,\qv)
    =\frac{1}{n}\left<\hat{\zv}-\hat{\xv},\hat{\zv}-\hat{\xv}\right>_\Id
    =\frac{1}{n}\sum_{i=1}^n (\hat{z_i}-\hat{x_i})^2.
\end{equation*}
This means that, \emph{minimizing the $\qv$-MSE is equivalent to minimizing the $\ell_2$ error energy in the IAGFT domain}. Based on this fact, we propose an image coding scheme that integrates IAGFT into the JPEG framework. The diagram is shown in Fig.~\ref{fig:diagrams}, where the values in $\Qm$ are quantized and transmitted as signaling overhead for the decoder to uniquely reconstruct the image. Further details for implementation will be described in Sec.~\ref{sec:experiments}.

Next we provide a pixel domain error analysis under uniform quantization noise assumption. Let $\bm{\varepsilon}_p$ and $\bm{\varepsilon}_t$ be the vectors of errors within a block in the pixel and IAGFT domain, respectively.  
Then, the variance of the $i$-th element in $\bm{\varepsilon}_p$ is
\begin{align*}
    %& \mathbb{E}\left[\varepsilon_p(i)\right]=\mathbb{E}\left[\ev_i^\top\Um\bm{\varepsilon}_t\right]=0, \\
    \mathbb{E}\left[\varepsilon_p(i)^2\right] = \mathbb{E}\left[(\ev_i^\top\Um\bm{\varepsilon}_t)^2\right]
    = \trace\left(\Um^\top\ev_i\ev_i^\top\Um \cdot  \mathbb{E}\left[\bm{\varepsilon}_t\bm{\varepsilon}_t^\top\right]\right),
\end{align*}
where $\ev_i$ is the $i$-th standard basis. Denote the quantization step size for the $j$-th transform coefficient as $\Delta_j$ and model the quantization noise with uniform distribution $\varepsilon_t(i)\sim\text{Unif}(-\Delta_i/2,\Delta_i/2)$. Thus, we have $\mathbb{E}\left[\bm{\varepsilon}_t\bm{\varepsilon}_t^\top\right]=\text{diag}(\Delta_1^2,\dots,\Delta_n^2)/12$. When a uniform quantizer with $\Delta_i=\Delta$ is used for all $i$, 
\begin{equation}
\label{eq:epi2}
  \mathbb{E}\left[\varepsilon_p(i)^2\right] 
  = \frac{\Delta^2}{12}\trace\left(\Um^\top\ev_i\ev_i^\top\Um\right)
  = \frac{\Delta^2}{12}\trace\left(\ev_i\ev_i^\top\Qm^{-1}\right) = \frac{\Delta^2}{12q_i},
\end{equation}
where we have used $\Um\Um^\top=\Qm^{-1}$, which follows from the fact that $\Um^\top(\Qm\Um)=\Id=(\Qm\Um)\Um^\top$. With \eqref{eq:epi2}, we know that the expected WMSE for this block is
\[ 
\mathbb{E}\left[\text{WMSE}(\zv,\xv,\qv)\right]
=\frac{1}{n}\sum_{i=1}^n q_i\left(\mathbb{E}\left[\varepsilon_p(i)^2\right]\right)=\frac{\Delta^2}{12},
\]
which only depends on the quantization step.

Note that the scheme shown in Fig.~\ref{fig:diagrams}(b) can be viewed as a bit allocation method. When $q_j=2q_i$ for pixels $i$ and $j$ within a block, the quantization error of IAGFT coefficients tends to contribute more error to pixel $j$ than to pixel $i$ in the pixel domain. Implicitly, this indicates that more bits are spent to accurately encode pixel $j$. On the other hand, if $\Qm_\ell=2\Qm_k$ for blocks $k$ and $\ell$, we can show that the IAGFT coefficients will satisfy $\xv_\ell=\sqrt{2}\xv_k$, meaning that the encoder tends to use more bits for block $\ell$ than for block $k$.

\subsection{SSIM-Driven Weights for WMSE}
\label{subsec:ssim_wmse}

In this work, we adopt the structural similarity (SSIM) as the target metric for perceptual quality, and design a WMSE to optimize it\footnote{In fact, our proposed method can be applied for any arbitrary WMSE. The application of this method based on other metrics such as Video Multimethod Assessment Fusion (VMAF) \cite{vmaf} is considered for future work.}. SSIM is one of the most popular image quality assessment metrics, with many experiments demonstrating its better alignment with perceptual visual quality as compared to MSE \cite{wang2004image}. For a distorted image $\zv$ and its reference image $\xv$, the definition of SSIM is
\begin{align}
    & \text{SSIM}(\xv,\zv) = \frac{1}{n}\sum_{i=1}^n \text{SSIM}(x_i,z_i),%\quad \text{where}
    \nonumber\\
    \label{eq:localssim}
    & \text{SSIM}(x_i,z_i) = \frac{2\mu_{x_i}\mu_{z_i}+c_1}{\mu_{x_i}^2+\mu_{z_i}^2+c_1} \cdot \frac{2\sigma_{x_i z_i}+c_2}{\sigma_{x_i}^2+\sigma_{z_i}^2+c_2},
\end{align}
where $\mu_{x_i}$ and $\sigma_{x_i}^2$ are local mean and variance around pixel $i$ and the summation is taken over all $n$ pixels in the image.

\begin{figure}
    \centering
    \includegraphics[width=.4\textwidth]{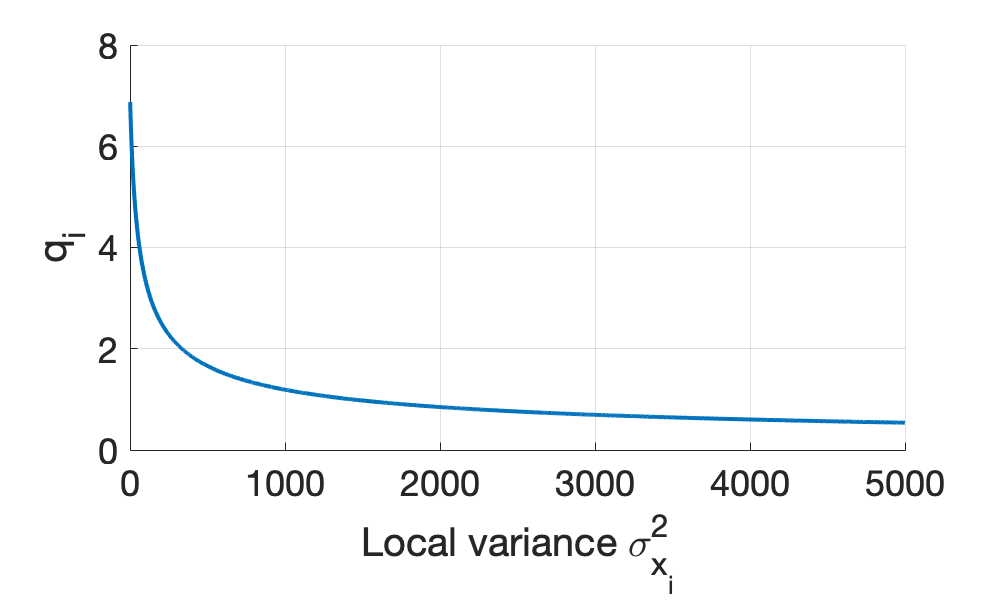}
    \caption{Values of $q_i$ with respect to local variance, with $\Delta=8$.}
    \label{fig:qi_vs_varx}
\end{figure}

We denote $\zv=\xv+\bm{\varepsilon}_p$, and assume that $\xv$ and $\bm{\varepsilon}_t$ are independent. Based on the statistics of $\bm{\varepsilon}_p$ derived in Sec.~\ref{subsec:coding_iagft}, we have $\mu_{z_i}=\mu_{x_i}$, $\sigma_{x_i z_i}^2 = \sigma_{x_i}^2$, and $\sigma_{z_i}^2=\sigma_{x_i}^2+\Delta^2/12q_i$. Thus, the local SSIM in \eqref{eq:localssim} reduces to
\[
    \text{SSIM}(x_i,z_i) = \frac{2\sigma_{x_i}^2+c_2}{2\sigma_{x_i}^2+c_2+\Delta^2/(12q_i)} = \frac{q_i}{q_i+\gamma_i}, .
\]
where $\gamma_i=\Delta^2/12(2\sigma_{x_i}^2+c_2)$.
To obtain $\qv$ that maximizes the SSIM, we introduce an upper bound for $\sum_i q_i$ as a proxy of the bitrate constraint, and solve
\begin{equation}
\label{eq:qi_problem}
    \underset{\qv}{\text{maximize}}\quad \frac{1}{n}\sum_{i=1}^n\frac{q_i}{q_i+\gamma_i} \qquad \text{subject to} \quad \sum_{i=1}^n q_i \leq n.
\end{equation}
It can be shown that this problem is convex in $\qv$. 
% We define the Lagrangian cost function as
% \[
%   \Jc(\qv,\lambda) = -\frac{1}{n}\sum_{i=1}^n\frac{q_i}{q_i+\gamma_i} + \lambda\left( \sum_{i=1}^n q_i - n \right),
% \]
% where $\lambda$ is the Lagrange multiplier. By setting the derivatives w.r.t $q_i$ and $\lambda$ to zero, we
Using the Lagrangian cost function and Karush–Kuhn–Tucker (KKT) conditions, we can obtain a closed-form solution:
\begin{equation}
\label{eq:q_closedform}
    q_i = \frac{(n+\sum_{i=1}^n \gamma_i)\sqrt{\gamma_i}}{\sum_{i=1}^n\sqrt{\gamma_i}} - \gamma_i.
\end{equation}

While $\qv$ is a high dimensional vector (with dimension $n$, the number of pixels in the image), \eqref{eq:q_closedform} provides an efficient way to obtain the optimal solution, which only depends on the quantization step and local variance. The computation of all $q_i$ can be carried out in $\Oc(n)$ time. Fig.~\ref{fig:qi_vs_varx} shows the resulting of $q_i$ with different local variance values and a fixed quantization step size. The fact that $q_i$ decreases with respect to local variance means that larger weights are used for pixels in uniform or smooth regions of the image, which in turn results in higher quality in those regions.

\section{IAGFT Transform Bases}
\label{subsec:bases}

\begin{figure}
    \centering
    % \subfigure[]{
    % \includegraphics[width=.23\textwidth]{figures/basis4_dct.png}}%
    \subfigure[]{
    \includegraphics[width=.47\textwidth]{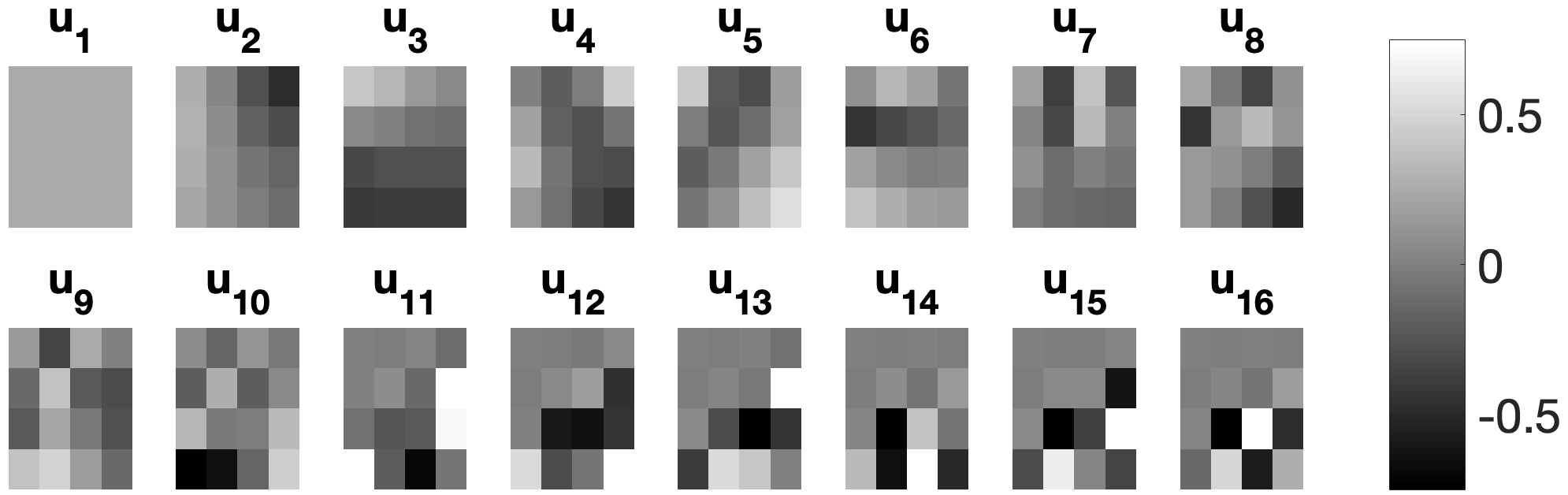}}\\
    \begin{minipage}[b]{0.47\textwidth}
       \subfigure[]{\includegraphics[width=.49\textwidth]{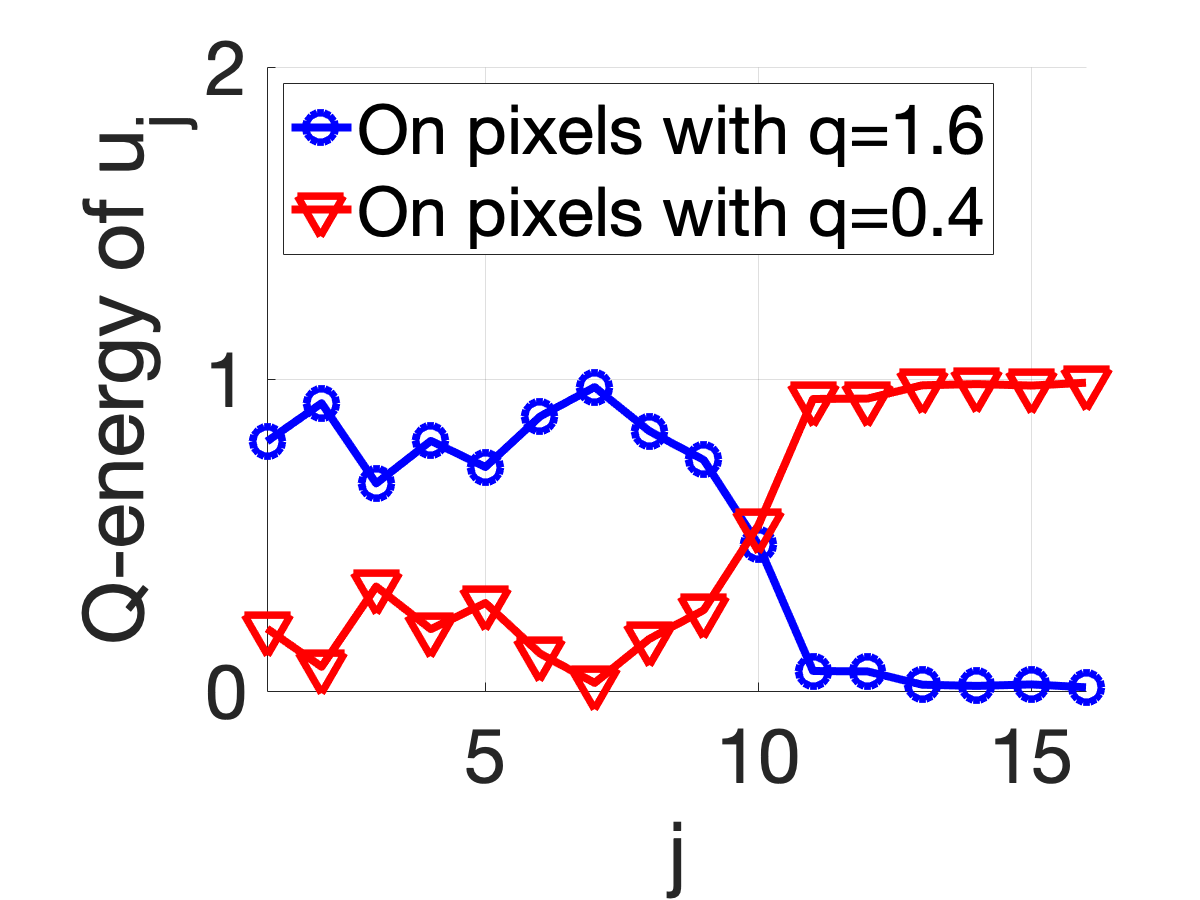}}%
       \subfigure[]{\includegraphics[width=.49\textwidth]{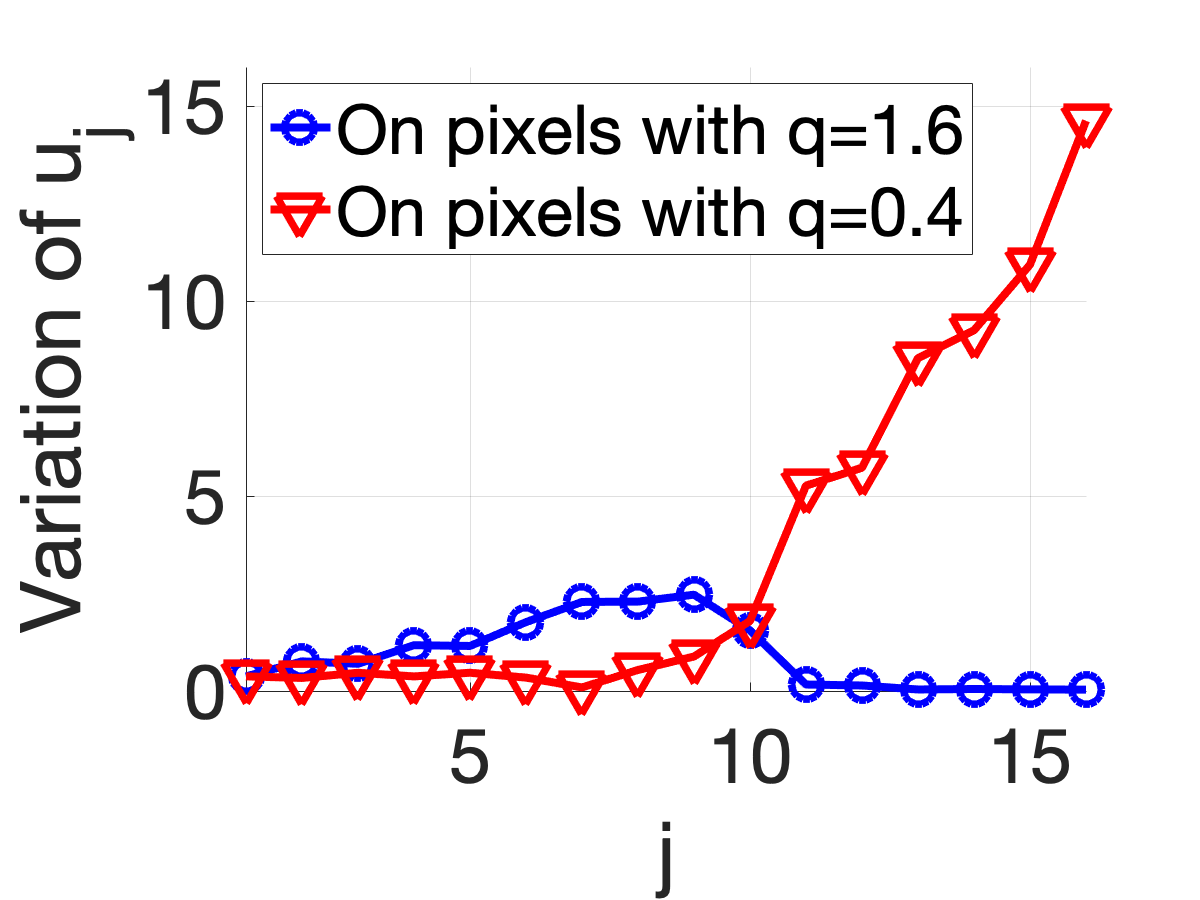}}
     \end{minipage}
    \caption{The (a) basis, (b) $\Qm_0$-energy, and (c) variantions of $\Qm_0$-IAGFT modes. In (b) and (c), two curves represent quantities for pixels with $q_i=1.6$ and with $q_i=0.4$, respectively.}
    \label{fig:iagft_bases}
\end{figure}

In Fig.~\ref{fig:iagft_bases}(a), we show the 2D $\Qm_0$-IAGFT basis, where $\Qm_0$ is the diagonal matrix associated to a $4\times 4$ block with WMSE weights:
\[
\begin{pmatrix} 1.6 & 1.6 & 1.6 & 1.6 \\ 1.6 & 1.6 & 1.6 & 0.4 \\ 1.6 & 0.4 & 0.4 & 0.4 \\ 0.4 & 0.4 & 0.4 & 0.4
\end{pmatrix}.
\]
Note that the pixels in the top left corner have larger weights, while the weights sum to 16 as in the $\Id$-IAGFT (i.e. DCT). In Fig.~\ref{fig:iagft_bases}(b)(c) we show the $\Qm_0$-energy and variation of each basis function, within top-left regions ($q_i=1.6$) and within bottom-right regions ($q_i=0.4$). In these figures we observe that those IAGFT basis functions corresponding to low to medium frequencies  (i.e., $\uv_1$ to $\uv_9$) have increasing variations for pixels in the top left region, while those corresponding to the highest frequencies ($\uv_{11}$ to $\uv_{16}$) are localized to the lower right corner. This means that in the pixel domain, the energy in the highly weighted area (with large $q_i$) will be mostly captured by low frequency IAGFT coefficients.

As a second example, we consider the case with $\Qm=k\Id$, where the IAGFT coefficients are $\sqrt{k}$ times the DCT coefficients. In this case, when we apply a $k\Id$-IAGFT followed by a uniform quantization with step size $\Delta$, it is equivalent to applying a DCT followed by a quantization with step size $\sqrt{k}\Delta$. Therefore, this special case reduces to a block-wise quantization step adjustment scheme, as in related work such as \cite{hontsch2002adaptive}. This means that, our scheme can be viewed as a generalization of quantization step adjustment method, while our method can adapt the quality per pixel, which is finer than a per block adaptation.

\section{Experimental Results}
\label{sec:experiments}

\begin{figure}[th]
    \centering
    \includegraphics[width=.48\textwidth]{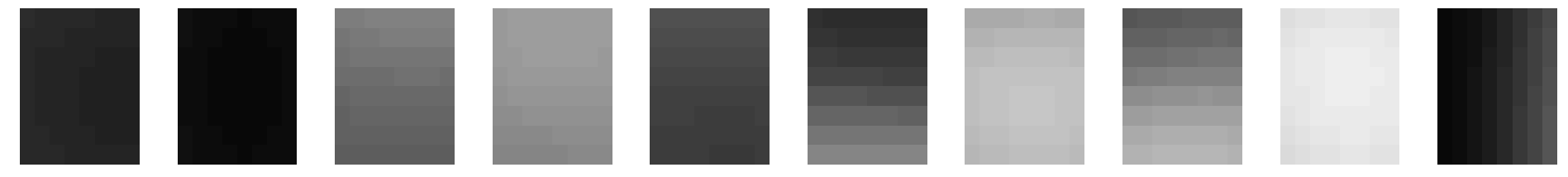}
    \caption{Vector quantization codewords of $q_i$ for $8\times 8$ blocks.}
    \label{fig:codewords}
\end{figure}

\begin{figure}[th]
    \centering
    \subfigure[]{
    \includegraphics[width=.23\textwidth]{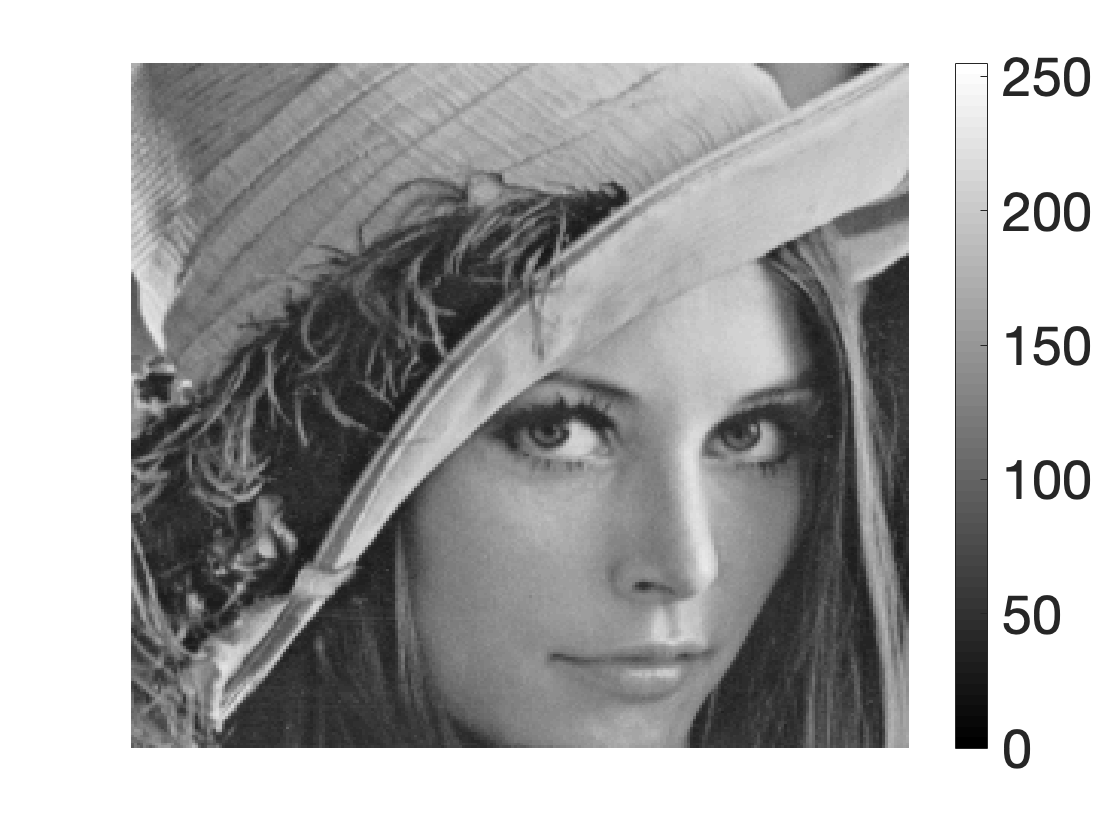}}%
    \subfigure[]{
    \includegraphics[width=.23\textwidth]{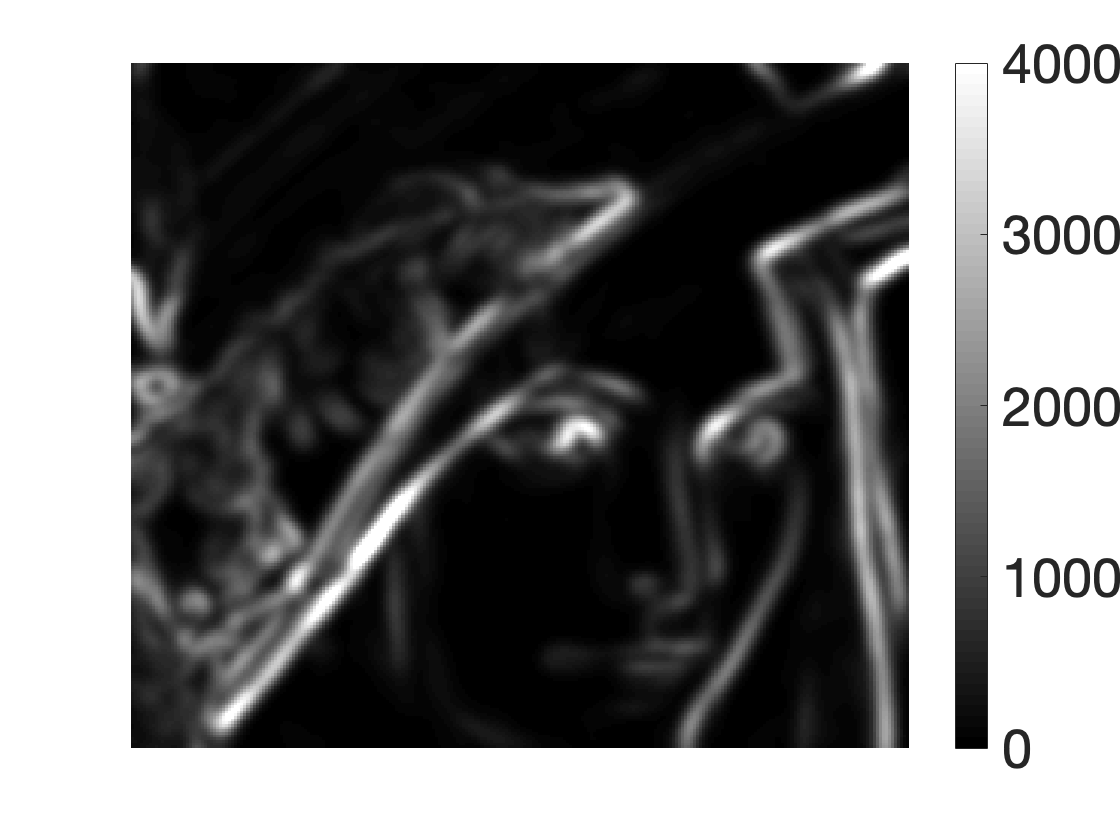}}\\
    \subfigure[]{
    \includegraphics[width=.23\textwidth]{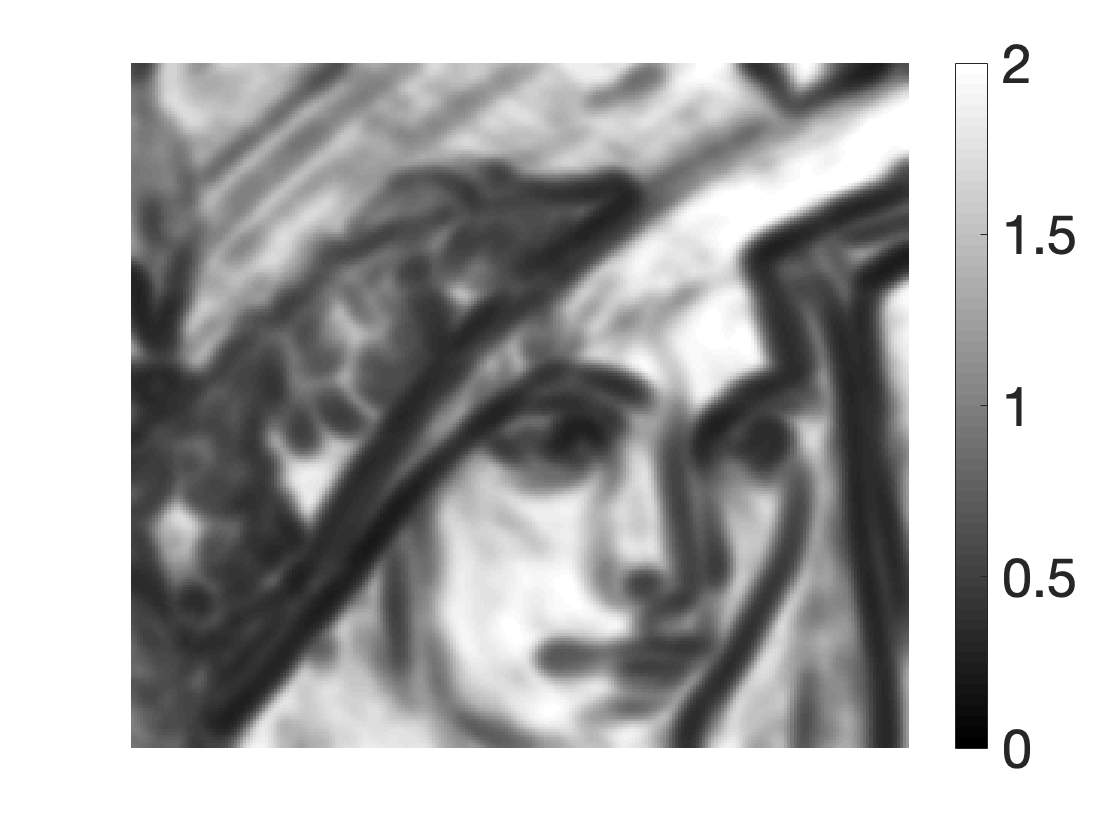}}%
    \subfigure[]{
    \includegraphics[width=.23\textwidth]{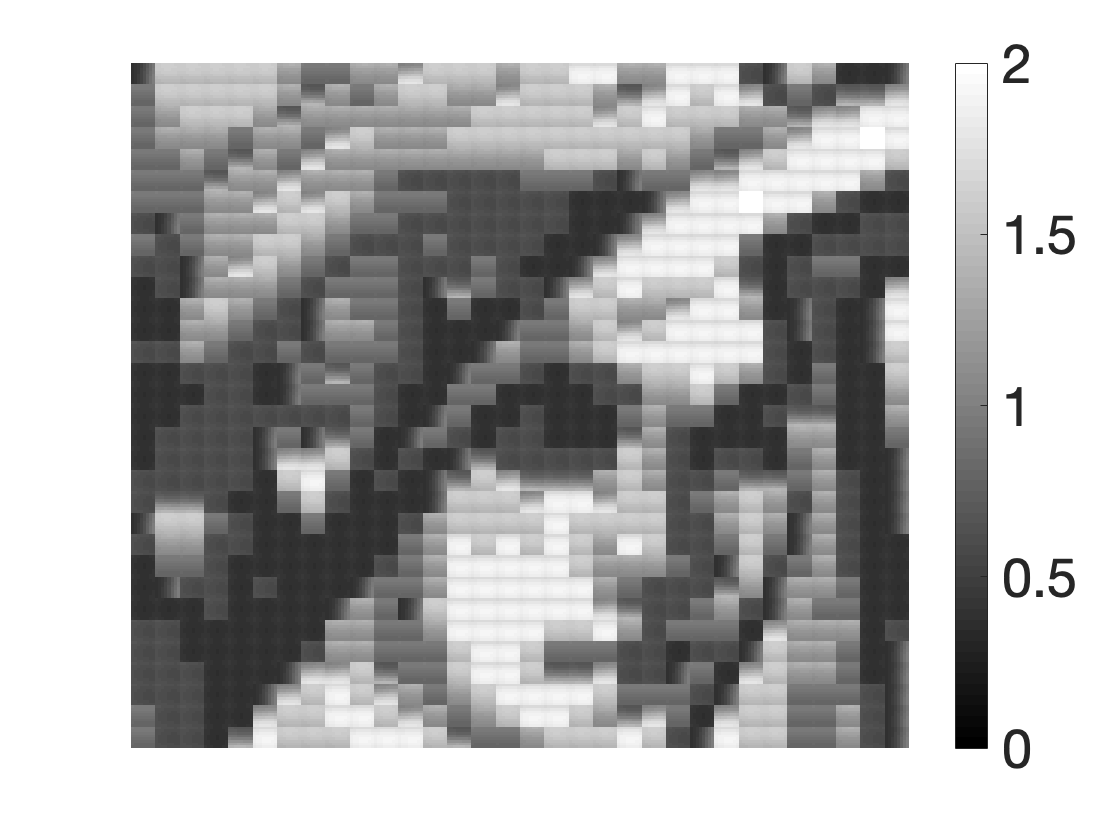}}
    \caption{An example: (a) original image, (b) local variance map (c) $q_i$ map, and (d) quantized $q_i$ map with vector quantization.}
    \label{fig:lena_example}
\end{figure}

To demonstrate the effectiveness of the proposed framework, we apply the coding scheme illustrated in Fig.~\ref{fig:diagrams} in JPEG. Note that the non-uniform quantization table in JPEG standard was designed based on perceptual criteria for DCT coefficients. We propose a similar non-uniform quantization for IAGFTs as follows. For an IAGFT with basis functions $\uv_k$, we find the unique representations in DCT domain, denoted as $\uv_k=\sum_{i=1}^n \phi_{ki}\vv_i$ with $\vv_i$ being the $i$-th DCT basis vector. Then, we choose quantization step associated to $\uv_k$ as a weighted mean: $\sum_{i=1}^n |\phi_{ki}^{(j)}|\Delta_i$, where $\Delta_i$ is the quantization step associated with $\vv_i$. In this way, when $\uv_k$ has low frequency in DCT domain, a small quantization step size will be used, and vice versa. The weights $\qv$ for WMSE are obtained from \eqref{eq:q_closedform}. The Laplacian of a uniform grid graph is used as variation operator to define the IAGFT.  
%A low-pass filter is then applied to the weights for robustness of performance. 
For signaling overhead, we apply a entropy-constrained vector quantization (VQ) \cite{gersho1992vector} with 10 codewords trained from the $8\times 8$ blocks of $q_i$ values in \texttt{house} image. The resulting codewords are shown in Fig.~\ref{fig:codewords}, and signaling overhead for each codeword is based on the selection frequency during VQ training. For each $8\times 8$ block of testing images, we quantize the corresponding $\qv$ to the closest codeword, and apply the associated $8\times 8$ non-separable IAGFT. We assume that transform bases and quantization tables for the IAGFTs represented by the codewords are embedded in the codec, so we do not require eigen-decomposition or side information for bases and quantizers. For illustration, Fig.~\ref{fig:lena_example}(b) shows the local variance map obtained as in SSIM formula \eqref{eq:localssim}, and Fig.~\ref{fig:lena_example}(c)(d) show that resulting $q_i$ obtained from \eqref{eq:q_closedform} and the quantized $q_i$ with VQ, respectively.

\begin{figure}[th]
    \centering
    % \subfigure[]{
    % \includegraphics[width=.23\textwidth]{figures/unif_psnr.png}}%
    % \subfigure[]{
    % \includegraphics[width=.23\textwidth]{figures/unif_ssim.png}}
    % \subfigure[]{
    % \includegraphics[width=.23\textwidth]{figures/qts_psnr.png}}%
    % \subfigure[]{
    % \includegraphics[width=.23\textwidth]{figures/qts_msssim.png}}
    \subfigure[]{
    \includegraphics[width=.23\textwidth]{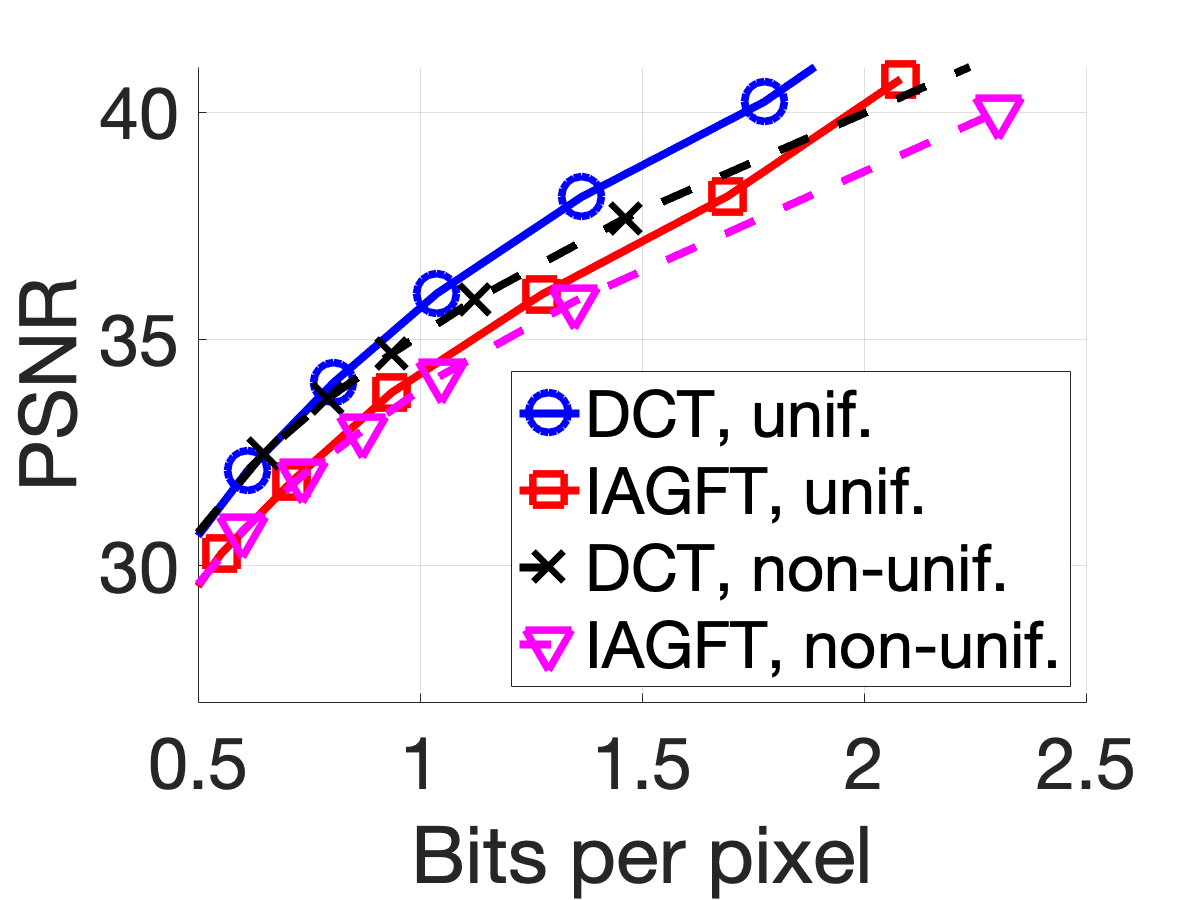}}%
    \subfigure[]{
    \includegraphics[width=.23\textwidth]{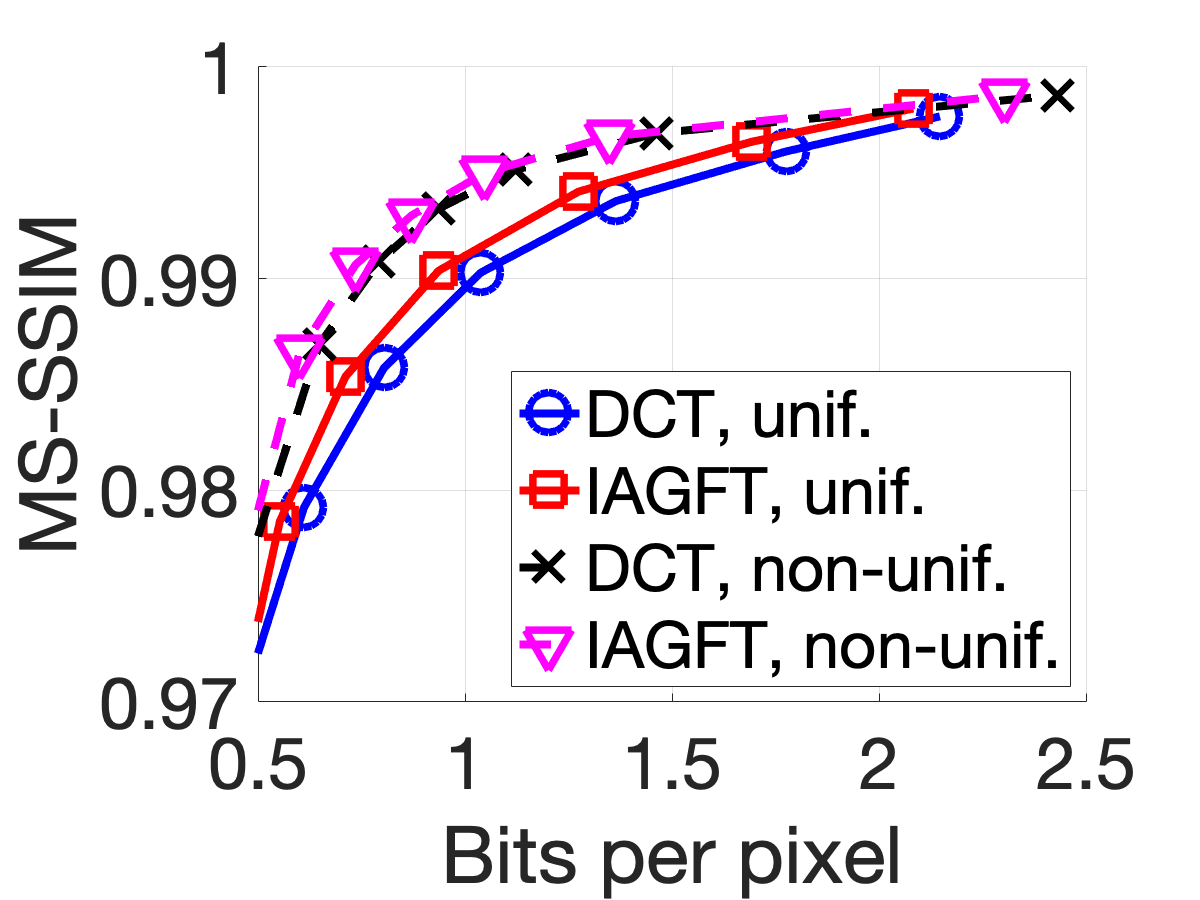}}
    \caption{RD curves for \texttt{Airplane} image in (a) PSNR and (b) MS-SSIM.}
    \label{fig:rdcurves}
\end{figure}

\begin{table}[t]
    \centering
    % \begin{tabular}{|cc|c|c|c|c|}
    % \hline
    %     & & Lena & Mandrill & Plane & Pepper \\
    % \hline
    % \multirow{3}{*}{U} & PSNR & 19.71\% & 12.80\% & 21.02\% & 17.47\% \\
    % & SSIM & 4.96\% & 2.03\% & 6.34\% & 5.96\% \\
    % & MS-SSIM & 0.39\% & -1.44\% & -1.20\% & 3.52\% \\
    % \hline
    % \multirow{3}{*}{NU} & PSNR & 17.08\% & 12.53\% & 20.43\% & 18.31\%  \\
    % & SSIM & 5.25\% & 3.34\% & 7.64\% & 6.17\% \\
    % & MS-SSIM & 2.36\% & 0.55\% & 0.95\% & 4.18\% \\
    % \hline
    % \end{tabular}
    \begin{tabular}{|c|c|c|c|c|c|}
    \hline
        \multicolumn{2}{|c|}{} & Mandrill & Lena & Airplane & Sailboat \\
    \hline
    \multirow{3}{*}{U} 
    & PSNR & 14.78\% & 17.64\% & 20.0\% & 17.69\% \\
    & SSIM & 1.28\% & 2.18\% & 3.81\% & 1.64\% \\
    & MS-SSIM & -2.09\% & -2.25\% & -8.18\% & -7.70\% \\
    \hline
    \multirow{3}{*}{NU} 
    & PSNR & 15.16\% & 14.45\% & 20.23\% & 16.43\%  \\
    & SSIM & 4.36\% & 3.66\% & 6.52\% & 5.70\% \\
    & MS-SSIM & -0.15\% & -0.93\% & -6.09\% & -0.98\% \\
    \hline
    \end{tabular}
    \caption{Bit rate reduction with respect to DCT-based schemes. Negative numbers correspond to compression gains. U and NU stand for uniform and non-uniform quantization tables, respectively.}
    \label{tab:rd}
\end{table}

The RD performances in terms of PSNR and SSIM are shown in Fig.~\ref{fig:rdcurves}. We observe that the proposed scheme leads to a loss in PSNR, while giving a compression gain in SSIM. In Table~\ref{tab:rd}, we show the BD rate performance for several benchmark images. 
The proposed scheme outperforms DCT in multi-scale SSIM (MS-SSIM) \cite{wang2003multiscale} for all test images. 
% We observe that this gain is more consistent when bit rate is higher than 2 bpps (not shown in the figures) for the following reasons:
% \begin{itemize}
%     \item The signaling overhead of 3 bits per block is relatively large for low bit rates.
%     \item The error model in Sec.~\ref{subsec:ssim_wmse} is derived based on the uniform distributed noise model, which is less accurate when the quantization step size is larger. As a result, the model in \eqref{eq:epi2} is more suitable for higher bit rates.
% \end{itemize}
Note that, while the formulation \eqref{eq:qi_problem} is based on a uniform quantization assumption, our method can also achieve a compression gain when a non-uniform quantization table is used. We also observe that for this experiment, the side information accounts for 6\% to 8\% of the bit rates. Thus, further optimization for signaling overhead may lead to a coding gain in SSIM.

\section{Conclusion}
\label{sec:conclusion}
In this work, we consider the weighted mean square error (WMSE) as an image quality metric, as a generalization of the widely used MSE. By selecting proper weights, the WMSE offers a high flexibility and enables us to better characterize human perceptual quality than the MSE. In order to optimize the WMSE-based image quality, we proposed a novel image coding scheme using irregularity-aware graph Fourier transform (IAGFT). Based on the generalized Parseval's theorem, we have shown the optimality in terms of WMSE when uniform quantization is performed in the IAGFT domain. We then design weights to maximize the structural similarity (SSIM), where the weights are determined by the local image variances and quantization step size. 
%When integrated into JPEG standard, our method with the associated SSIM-driven WMSE provides a compression gain in SSIM. 
When integrated into JPEG standard, our method with the associated SSIM-driven WMSE can provide a compression gain in MS-SSIM. 
In the future, we will extend this method to schemes with non-uniform quantization and consider the integration into existing video codecs.

%\vfill\pagebreak

\bibliographystyle{IEEEbib}
\bibliography{refs}

% the version with appendix (proof of the error bound)
\vfill\pagebreak

\end{document}

%% file: macros.tex
\usepackage{bm}

\long\def\comment#1{}

\newfont{\bbb}{msbm10 scaled 700}

\newfont{\bb}{msbm10 scaled 1100}

\newcommand{\ev}{{\bf e}}

\newcommand{\qv}{{\bf q}}

\newcommand{\uv}{{\bf u}}

\newcommand{\vv}{{\bf v}}
\newcommand{\xv}{{\bf x}}
\newcommand{\yv}{{\bf y}}
\newcommand{\zv}{{\bf z}}
\newcommand{\zerov}{{\bf 0}}
\newcommand{\onev}{{\bf 1}}

\newcommand{\Dm}{{\bf D}}

\newcommand{\Fm}{{\bf F}}

\newcommand{\Id}{{\bf I}}

\newcommand{\Lm}{{\bf L}}

\newcommand{\Qm}{{\bf Q}}

\newcommand{\Um}{{\bf U}}
\newcommand{\Wm}{{\bf W}}

\newcommand{\Ec}{{\cal E}}

\newcommand{\Gc}{{\cal G}}

\newcommand{\Nc}{{\cal N}}
\newcommand{\Oc}{{\cal O}}

\newcommand{\Vc}{{\cal V}}

\newcommand{\Lambdam}{\hbox{\boldmath$\Lambda$}}

\newcommand{\trace}{{\hbox{tr}}}

\newtheorem{definition}{Definition}